\documentclass[12pt,preprint]{aastex}

\shorttitle{The Absorption and Emission Kinematics in the Mg~II Absorber 
Towards Q1331+17}
\shortauthors{Ellison, Mallen-Ornelas \& Sawicki}

\begin{document}


\title{The Absorption and Emission Kinematics
in the $z=0.7450$ Mg~II Absorber Towards Q1331+17\footnote{This paper 
is based on observations collected with the VLT at the European Southern
Observatory, Chile, as part of ESO programs 67.A-0334 and
68.A-0170.}.}


\author{Sara L. Ellison}
\affil{European Southern Observatory, Casilla 19001, Santiago, Chile}
\affil{P. Universidad Cat\'olica de Chile, Casilla 306, Santiago 22, Chile}
\email{sellison@astro.puc.cl}

\and

\author{Gabriela Mall\'en-Ornelas\altaffilmark{2,3}}
\affil{Princeton University Observatory, Peyton Hall, Princeton, NJ 08544, USA}
\affil{P. Universidad Cat\'olica de Chile, Casilla 306, Santiago 22, Chile}
\email{mallen@astro.princeton.edu}

\and

\author{Marcin Sawicki\altaffilmark{2}} 
\affil{
Dominion Astrophysical Observatory, 
Herzberg Institute of Astrophysics,
National Research Council, 
5071~West Saanich Road, 
Victoria, B.C., V9E 2E7, Canada.
}
\email{marcin.sawicki@nrc.ca}

\altaffiltext{2}{Guest User, Canadian Astronomy Data Centre, which
is operated by the Herzberg Institute of Astrophysics,
National Research Council of Canada}
\altaffiltext{3}{Current address: Harvard-Smithsonian Center for Astrophysics,
MS 15, 60 Garden Street,
Cambridge MA 02138, USA}

\begin{abstract}

We present a comparative analysis of the galaxy emission and QSO absorption
kinematics of a $z_{abs} \sim 0.7450$ Mg~II system and its candidate
absorbing galaxy (G5) located 3.86 arcsecs (28.3 $h_{70}^{-1}$ kpc)
from the QSO.  We have obtained a spectrum of the galaxy
candidate, previously identified as a luminous edge-on disk, and
detect the [OII]~$\lambda 3727$~\AA\ doublet at a systemic redshift of
$z_{sys}=0.7450$.  From slit spectroscopy of this
galaxy, we find $v_{rot} \gtrsim 210$ km/s, and possibly as
large as 350 km/s.  Plotted on the same velocity scale, the systemic
redshift of the galaxy coincides with the center of
the absorption system, although the absorption components span 
more than 100 km/s in either direction.  However, once the sense of 
the rotation is taken into account, there is no absorbing
gas at the projected velocity of the disk rotation curve.
This implies incompatibility with a simple disk scenario.
Moreover, a re-analysis of archival HST data reveals that the galaxy is only
0.3 $L_{\star}$, considerably less luminous than previously reported
in the literature.  This is incompatible with the established
Tully-Fisher relation at this redshift, unless approximately 2
magnitudes of total extinction are invoked.  Careful inspection of the
archival HST data reveal that G5 may well be composed of two
galaxies, although the quality of the data do not permit a
detailed investigation of this.  This possibility is further supported by the
identification of a second faint emission line at $\lambda_{obs}=5674$ \AA, 
whose distinct
spatial and velocity profile indicates that it arises in a different
galaxy at a different redshift.    Analysis of the
absorption lines shows evidence for superbubbles in the interstellar
medium (ISM) of the
absorbing galaxy, based on the striking symmetry between components
and large Mg~I/Mg~II and Mg~I/Fe~II ratios, indicative of large 
densities.  The large velocity separations between line pairings,
$\Delta v \sim 150$ km/s indicates that these bubbles may be powered
by OB associations comparable to the largest observed at $z=0$ and that the
gas is probably enriched to at least 1/10 solar metallicity.  This is
consistent with observations at low redshift that extended Mg~II halos
are often seen in galaxies that contain disturbed gas.
Superbubbles may also explain why the absorber has a relatively large
Mg~II equivalent width relative to the luminosity of the associated
galaxy (or galaxies).

\end{abstract}


\keywords{quasars: absorption lines --- galaxies:kinematics and
dynamics --- galaxies: halos --- galaxies: evolution --- ISM:bubbles
--- quasars:individual (Q1331+17)}


\section{Introduction}

Quasar absorption lines have developed into a key tool for probing the
evolving population of galaxies from high redshift to the present day.
Since selecting galaxies based on their absorption, rather than
emission properties circumvents selection effects associated with
Malmquist bias, this is a potentially powerful technique for tracing
the evolution of gas repositories over the bulk of the Hubble time.
One of the most challenging aspects of quasar absorption line research
is the association between different types of absorbers and their
galactic counterparts; without this crucial information we are lacking
an infra-structure within which to interpret the results.  This crux
is perhaps best highlighted by the work which has focussed on Damped
Lyman Alpha systems (DLAs).  DLAs have been variously associated with
galactic disks (Prochaska \& Wolfe 1997, 1999), low surface brightness
galaxies (Jimenez et al.\ 1999), dwarfs (Nulsen et al.\ 1998), merging
protogalactic clumps (Haehnelt et al.\ 1998) and galactic winds (Schaye
2001).  A key diagnostic to the nature of high redshift DLAs is
the combined study of low redshift absorbers whose counterpart
galaxies can be investigated in tandem.  Establishing an understanding 
of low redshift galaxy absorption properties through the profiling
of abundances, luminosities, morphologies and star formation histories
is essential for extrapolation to high redshifts where only absorption
line abundances are available.
Unfortunately, the direct identification of DLA
galaxies has seen relatively slow progress.  Some 20 DLAs now have
tentative counterparts (e.g., Le Brun et al.\ 1997; Turnshek et al.\ 2001;
Nestor et al.\ 2001; Bowen, Tripp \& Jenkins 2001), although only 
approximately 25\%
of these have actually been confirmed spectroscopically.  Clearly,
increasing the number of spectroscopic confirmations is a priority,
one which has been highlighted by the results of Rosenberg \&
Schneider (2002) who warn against simply assigning the
identification to the nearest bright galaxy.

The characterisation of strong Mg~II absorbers at $z < 1.5$ has been
significantly more successful than for classical DLAs, with the identification
of a galaxy candidate within 40 kpc in almost every case (Steidel,
Dickinson \& Persson 1994; Steidel et al.\ 1997), many
of which have been spectrscopically confirmed.  Modelling of the
absorption line profiles of Mg~II systems has been undertaken in a
similar vein to DLAs (e.g., Lanzetta \& Bowen 1992; Charlton \&
Churchill 1998; Bond et al.\ 2001b), including models of radial infall
to explain the almost unit covering factor (but see also Charlton \&
Churchill 1996).  Most recently, Steidel et
al. (2002) have directly investigated the connection between absorption
and emission kinematics in 5 Mg~II galaxies and found that the
absorption profiles seem to be consistent with rotation, although the
kinematics cannot be explained by a simple disk model.  To our
knowledge, the kinematics of absorber-galaxy systems have
been investigated in only 2 other studies (Barcons,
Lanzetta, \& Webb 1995; Lanzetta et al.\ 1997).  Although the former
study also finds kinematics consistent with disk rotation, Lanzetta
et al (1997) intriguingly find the absorption gas moving counter to the
rotation direction of the $0.4 L_{\star}$ galaxy disk.

Here we present the results of a multi-faceted investigation of the
kinematics of a Mg~II absorber at $z \sim 0.745$ towards Q1331+17 which
combines high resolution QSO absorption line data with slit
spectroscopy and HST imaging of the galaxy counterpart.  This absorber
not only represents an eighth system for which the connection between
absorption and emission kinematics can be explored at intermediate
redshift, but additionally
allows us to investigate the presence of superbubbles recently
suggested by Bond et al.\ (2001a).  Bond et al.\ (2001b) have proposed
that a subset of strong Mg~II absorbers may be associated with
superbubbles or winds; shells of expanding gas scooped up in the ISM
of absorbing galaxies (e.g. Bond et al. 2001a; Rauch et al. 2002).  
The $z_{abs} \sim 0.745$ Mg~II system towards Q1331+17 has been
previously investigated by Bond et al.\ (2001a) as a particular case study 
for bubbles due to the striking pair-wise structure of absorption lines
seen in Keck HIRES spectra, first highlighted by Churchill, Vogt \&
Steidel (1995).  Large-scale outflows associated with such
winds have been observed in a variety of environments over a range of
redshifts, such as Lyman break galaxies 
(Pettini et al.\ 2001, 2002), high redshift
lensed galaxies (Frye, Broadhurst and Benitez 2002) and local
starburst galaxies (Heckman et al.\ 2000).  The existence of such
superbubbles that burst out of the galaxy may be a significant source
of metal enrichment in the intergalactic medium 
(Giroux \& Shull 1997; Ellison et al.\ 2000; Ferrara et al.\
2001).  The UVES data presented here, although
of only slightly superior resolution than the HIRES data used by Bond
et al.\ (2001a), reveal even more strikingly the pairwise symmetries
of the absorption features, which we discuss in the context of
superbubbles.

The paper is organized as follows.  In \S2 we describe multi-object
spectroscopy of candidate galaxy counterparts, echelle spectroscopy of
the QSO itself and the re-analysis of archival WFPC2 data.
Determination of the redshift of galaxy G5 and a compilation of
absorption line profiles are presented in \S3.  In \S4, we discuss the
connection between absorption and emission kinematics and the
implication for the superbubble scenario.

We adopt $\Omega_M = 0.3, \Omega_{\Lambda} = 0.7, H_0 = 70$ km/s/Mpc,
unless otherwise stated.

\section{Observations and Data Reduction}

Q1331+17 is a quasar well known to purveyors of absorption line
systems, since it is a bright ($V=16.7$) background target with a
$z_{abs} = 1.776$ DLA which has been widely studied in the literature
(e.g., Prochaska \& Wolfe 1999; Pettini et al.\ 1997; Lu et al.\ 1998).
In addition, there is a lower redshift ($z_{abs} = 0.745$) Mg~II
system which has been provisionally identified with an edge-on
$\sim$L$_{\star}$ disk at an impact parameter of 3.86 arcsec from the
QSO (Le Brun et al.\ 1997; Bond et al.\ 2001a) based on HST imaging.
There is also a Mg~II system at $z_{abs} = 1.328$ (Steidel \&
Sargent 1992) and a C~IV system at $z_{abs} = 1.446$ (Sargent, Steidel
\& Boksenberg 1988).  Finally, we have identified a new
C~IV system at $z_{abs}=1.374$ in the UVES data presented in \S2.2.

We have obtained ground-based spectroscopic observations of galaxies
in the QSO field in order to confirm the absorber identification for
the $z_{abs} = 0.745$ Mg~II system and study its internal kinematics.
In addition to the galaxy spectra, we have obtained high resolution
spectra of the QSO to study the absorber's metal lines and compare
their kinematics with those of the associated galaxy.

\subsection{Intermediate Resolution Multi-Object Spectroscopy of Galaxy 
Candidates}

\begin{figure}
\plotone{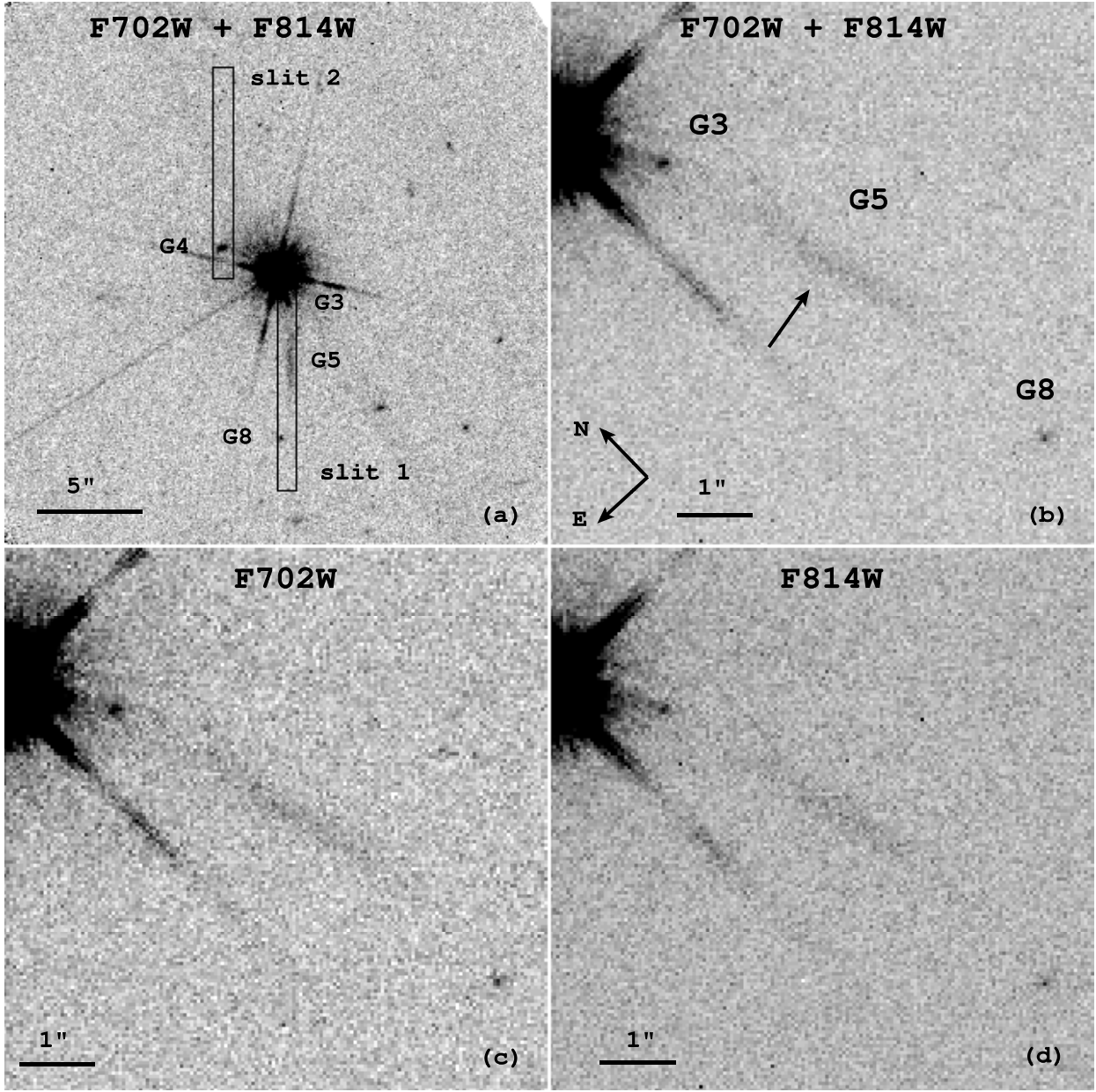}
\caption{Archival HST WFPC2 images, re-reduced as described in the text.
Panel (a) shows a rotated, combined image of the F702W and F814W filters
with the position of the central FORS slits superimposed on
candidate absorption galaxies.  Panel (b) shows a close-up view of
galaxy G5 in the combined filters.  The arrow indicates a possible 
`kink' in an apparently edge-on disk.  Panels (c) and (d) show
the same field of view for the individual filters.\label{fors_mask}}
\end{figure}

A total of 292 minutes of FORS2 service-mode spectroscopic
observations were carried out on 18 May 2001, 14 June 2001, 16 June
2001, and 14 July 2001.  Specifically, five 51-minute exposures and
one 37-minute exposure were taken for the two slits adjacent to the
QSO.  Grism 600RI+19 and 1$''$ slits were used, resulting in a
dispersion of 1.30~\AA/pix and average resolution of R$\sim$1200.  For
slits near the QSO, the wavelength range covered was 5500-8150~\AA.
Figure \ref{fors_mask}a shows an outline of the two central slits on a
rotated version of the HST WFPC2 archival image of the field.  For
consistency with earlier work, we have adopted the galaxy names used 
by Le Brun et al.\
(1997) and Bond et al.\ (2001a).  The end of the top slit covered G4,
and the bottom slit covered G3 and G5, the latter of which has been
previously identified by Le Brun et al.\ (1997) as the preferred 
absorbing galaxy candidate.  G5 is located 3.86 arcsecs from the QSO
which translates to 28.3 $h^{-1}$ kpc in our adopted cosmology.
Note that the fine alignment of these two slits near the QSO was done
with the aid of the WFPC image, which was re-sampled and superimposed
on the FORS2 image used for mask design.


The data were reduced using IRAF\footnote{IRAF is distributed by the
National Optical Astronomy Observatories, which is operated by the
Association of Universities for Research in Astronomy, Inc. (AURA)
under cooperative agrement with the National Science Foundation}.  A
bias frame was first subtracted from each spectroscopic exposure.
There were 1--2 pixel offsets in the object positions along the slit
in different frames, as well as in the position of the slit on the
CCD.  To account for these shifts, the multi-slit spectra were
registered so that the objects were positioned in the same pixel along
the spatial direction for every frame.  Subimages corresponding to 
individual slits were extracted and then reduced individually.  
Two-dimensional
wavelength-calibrated spectra from each slit and exposure were created
using IRAF tasks $identify, fitcoords$, and $transform$.  The
wavelength calibration was first determined based on a HeNeAr
comparison spectrum, and a second-order correction around the region
of interest was found by comparing the wavelength of sky lines in the
spectrum with theoretical values listed in Osterbrock et al.\
(2000)\footnote{Sky spectrum atlas available online at
http://www-mpl.sri.com/NVAO/download/Osterbrock.html.} available from
the VizieR database (Ochsenbein, Bauer, \& Marcout 2000).  Once the
wavelength solution was determined, the sky lines were subtracted
interactively from the raw spectra using the IRAF task $background$.
Cosmic rays were tagged on the sky-subtracted spectra using a
sigma-clipping algorithm, and replaced via linear interpolation of
neighbouring pixels (a necessary step before wavelength re-sampling).
The sky-subtracted and cosmic-ray free spectra were re-sampled into
two-dimensional wavelength-calibrated spectra, using the
previously-determined wavelength solution.  The individual
wavelength-calibrated and sky-subtracted spectra were co-added using
IRAF task $imcombine$ with cosmic-ray pixels masked to produce the
final two-dimensional spectra.

No emission lines or continuum were detected in G3 or G4, which is not
surprising due to their proximity to the QSO and faint magnitudes
($m(AB)_{814}=25.29$ and 24.47 from our re-analysis of the HST data,
see section 2.3).  However, G5 shows emission lines at 6505~\AA\ and
5674~\AA\ (see Figure~\ref{em_mos}).  A faint trace of continuum from
G5 is detected redwards of 6600~\AA\ (not seen in Figure~\ref{em_mos} due to
the limited wavelength range).  The identification of these lines
is discussed in \S~\ref{gal_conf} and summarized in Table~\ref{line_ids}.

\subsection{High Resolution Spectroscopy of Q1331+17}

Two hours of UVES observations of Q1331+17 were obtained on February
17 and 23, 2002 in service mode.  The 390+564 dichroic setting was
employed in order to give almost complete wavelength coverage from
3300 to 6650~\AA\ with small gaps between the 3 CCDs.  The spectra
were extracted using a customised version of the UVES pipeline
(Ballester et al.\ 2000) following the steps outlined in Ellison, Ryan
and Prochaska (2001).  By using a 1 arcsecond slit (aligned at the
parallactic angle) combined with 2$\times$2 binning, an average
resolution of $R \sim 42000$ was achieved.  Once converted to a linear
vacuum heliocentric scale, the two individual exposures were combined
with a weight proportional to the S/N (the two spectra were taken
under slightly different seeing and moon conditions) and normalised by
dividing through by a smoothed continuum.  The final S/N in the
combined spectrum varies between approximately 40 (per pixel) in the
blue and 70 in the red.

\subsection{Re-Analysis of HST WFPC2 Data}

We have re-reduced the archival HST WFPC2 data of Q1331+17 to re-assess the
absolute magnitude of G5 and its error estimate, since there are
various (discrepant) values available in the literature.  The archival
WFPC2 data were obtained through the CADC and consisted of six 600s F814W
images, as well as two 900s F702W images and two 600s F702W images.
Individual exposures were cleaned from cosmic rays by comparing
different exposures of the same mask using PPP (Yee, Ellingson \&
Carlberg 1996).  Additional cosmic ray residuals were tagged and
masked by hand.  Single images in F814W and F702W were produced by
co-adding the individual exposures in each filter.  All images in the
F814W filter were used.  For the F702W filter only the last three
images were co-added, since the first F702 exposure had a 
large number of cosmic rays covering galaxy G5.  
We measured the counts in a 0.8 $\times$ 3.2 arcsecond
rectangular aperture  which was centred on G5 and whose
long dimension was aligned with the major axis of the object.  The sky
level was estimated in similar rectangular-shaped apertures placed to
the east and west of G5 at the same distance from the QSO as the G5
aperture, but adjusted in width to avoid contamination by light from
the QSO diffraction spikes.  We also experimented with an alternative  
method of sky
subtraction which was to rotate the image 180$^{\circ}$ around the QSO and
then subtract it from the original image.  Both methods of sky
subtraction produced consistent results.  In addition to G5, aperture 
counts were also measured for a
number of galaxies in the field and converted to 
apparent magnitudes using the published STScI instrumental zero points.  
For G5, we determine (on the AB magnitude
scale) $m(814) = 23.26 \pm 0.3$ and $m(702) = 23.40 \pm 0.3$
(c.f. $m(814) = 21.47$ and $m(702) = 21.40$ from Le Brun et al.\ 1997),    
where the uncertainties include both the Poissonian count statistics and an
estimate of the systematics in aperture choice and sky subtraction.
This is in good agreement with the magnitude of G5 derived from 
ground based observations ($\cal R$ = 23.25 Steidel, private communication),
but differs by almost 2 magnitudes from the value of Le Brun et al. (1996).
The magnitudes of other measured galaxies in the field are consistent with
those of Le Brun et al. (1996).  We applied a K-correction appropriate for
an Scd galaxy (Coleman, Wu \& Weedman 1980), but this has a negligible
effect since the $I814$ band at the redshift of G5 corresponds
nearly identically to the rest-frame B-band.  Therefore, converting 
to our adopted cosmology yields $M_B(AB) = -19.35 \pm 0.3$.
Adopting $M_B^{\star}=-20.75$ (Efstathiou, Ellis \& Peterson 1998,
adjusted to our cosmology and magnitude system) this equates to $\sim
0.3L_{\star}$ compared with the value of $L \sim 1.4 L_{\star}$ determined
(for our cosmology) from the Le Brun et al.\ (1997) magnitude.  

\section{Results}

\subsection{Emission Line Identification and Galaxy Confirmation}
\label{gal_conf}

\begin{figure}
\plotone{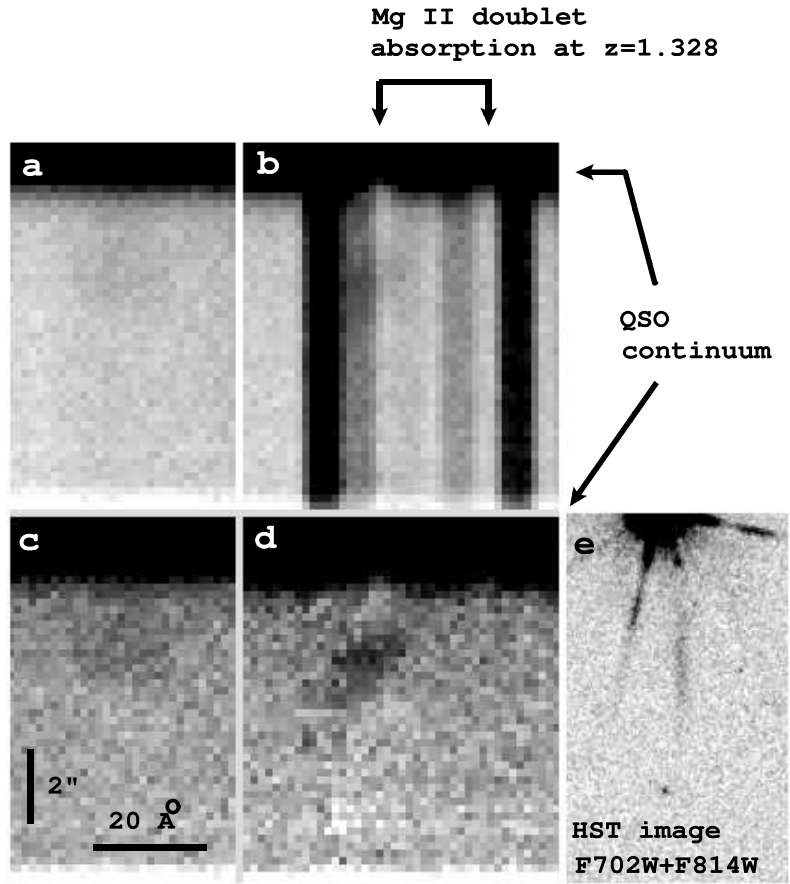}
\caption{Panels (a) and (b) show two postage stamps from the reduced 
2-D spectrum from the slit covering G5 (see Figure \ref{fors_mask}).  
The wavelength ranges are 5654.58 -- 5693.88 \AA\ and 6485.12 -- 6540.14
\AA\ for panels (a) and (b) respectively.  The QSO continuum appears
at the top of the slit and the Mg~II absorption system at $z=1.328$ is
marked.  The strong sky lines largely obscur the blue end of the
emission line at
6505 \AA.  Panels (c) and (d) have the sky background removed.  The
emission lines are now much clearer, but the lower spatial pixels
for the 6505 \AA\ line [panel (d)] 
may still have some significant residuals.  Panel (e) shows a 
scaled version of the WFPC2 image.  The emission line at 6505 \AA\
clearly matches the spatial extent of G5, whereas the line at 5674 \AA\
does not.
\label{em_mos}}
\end{figure}

\begin{figure}
\plotone{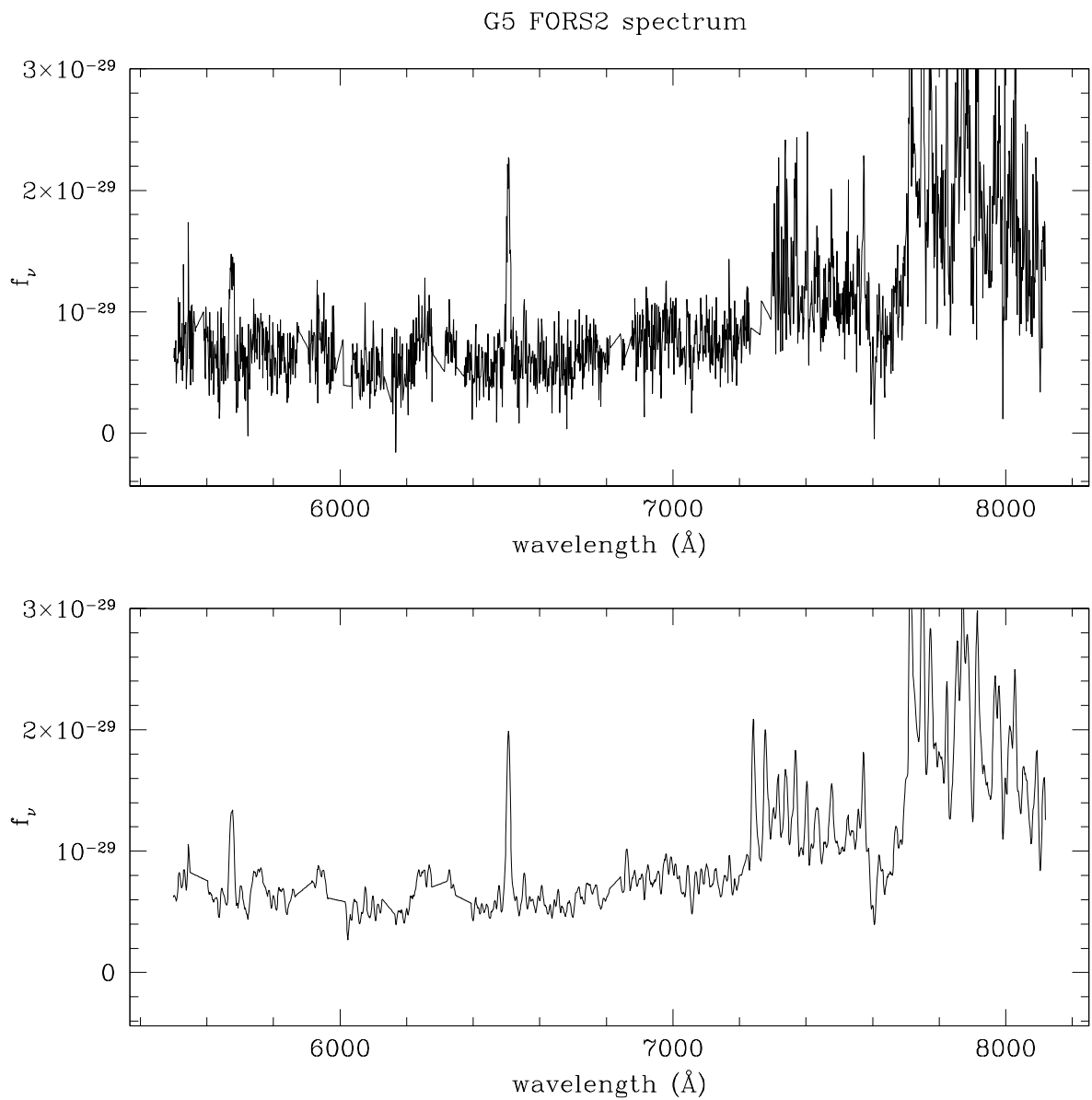}
\caption{1D  FORS2 spectrum of G5 in units of $F_{\nu}$ (although
the sky subtraction is quite uncertain, particularly given the
proximity to the QSO).  The top panel shows the spectrum at its
original resolution and bottom panel shows a smoothed version.  In
both cases, bright sky line residuals have been interpolated over for
clarity.  Note, however, the numerous sky emission residuals 
redward of 7200 \AA, and the telluric O$_2$ A band absorption 
redward of 7600 \AA.
The two emission lines at 5674 and 6505 \AA\ are clearly visible.\label{1d}}
\end{figure}

Two emission lines and a trace of faint continuum were found in the
spectrum of G5 (Figures \ref{em_mos} and \ref{1d}).  
There is a prominent line with a
noticeable velocity gradient across the galaxy at $\lambda_{obs} \sim
6505$~\AA\ (Figure~\ref{em_mos}b,d), as would be expected from a large
rotating disk galaxy.  A second, broader and fainter line is found at
$\lambda_{obs} \sim 5674$~\AA\ (Figure~\ref{em_mos}a,c).  This second
line is also slightly spatially displaced from the line at 6505~\AA\ (a 
\textit{conservative}
estimate of the offset between the line centers is 2 pixels), suggestive 
that the two lines may not be coming from the same physical location,
although the spatial offset is a marginal one.
More convincingly, there is a complete lack of kinematic structure in 
the 5674~\AA\ line when cross-correlated with a line template (see \S3.2),
whereas the 6505 \AA\ line shows a clear rotation curve.

There are 4 previously known absorbers in the line of sight to Q1331+17
(see Le Brun et al.\ 1997 for a summary), therefore
there are several possible galaxies which may be causing the emission lines.
Furthermore, we have searched our UVES spectrum (which has extended 
blue coverage)
for additional absorbers and identify one new C~IV system at 
$z_{abs} = 1.3742$.
Table~\ref{line_ids} lists the rest-frame wavelengths that each of the
two emission lines would have at the redshift of the four
absorbers known in the spectrum of Q1331+170.  These rest-frame
wavelengths were then compared with the lists of main emission lines
from nebular objects in Kinney et al.\ (1993) and Osterbrock (1989).
We now discuss each emission line in turn and present arguments
for the line identification in each case.

{\it (a) The emission line at 6505~\AA.}  There is a very good
wavelength match for the 6505~\AA\ line with [OII]~$\lambda$3727~\AA\ at $z =
0.7450$, in agreement with the galaxy identification for this system
given by Le Brun et al.\ (1997) and Bond et al.\ (2001a).  At the
moderate resolution of our spectra (R $\sim 1200$) we do not expect
the [OII] $\lambda$3727~\AA\ doublet ($\lambda\lambda$ 3726.032, 3728.815~\AA,
$\Delta \lambda/\lambda = 1340$) to be resolved.  However,
a comparison with sky lines in the spectrum shows that this emission
line has a broader profile than would be expected from a single unresolved
line.  A simple simulation shows that the width of the emission line at 
6505 \AA\ is in fact consistent with an unresolved 
[OII]~$\lambda$3727~\AA\ doublet.  

Coincidentally, there is a Mg~II absorption system at $z_{abs} = 1.328$
which puts Mg~II $\lambda$2796 at precisely the same observed wavelength as
[OII]~$\lambda$3727~\AA\ at $z \sim 0.7450$.    
The absorption doublet can in fact be
seen in the QSO continuum in Figure~\ref{em_mos}, with the bluer 
component lined up with the
emission line.  However, there is no sign of emission at the
wavelength that would correspond to the 2803~\AA\ component of the
Mg~II doublet.  Although the $f$-value ratio of MgII 2796:2803 is 2:1,
our spectrum is sufficiently sensitive that we would have resolved and
detected the
redder doublet feature were it present. Furthermore, galaxies generally 
show Mg~II in absorption rather than emission (e.g., Kinney et al.\ 1993).  
Thus, we rule out Mg~II at $z=1.328$ as the cause for the emission line
(see also the discussion on the emission line at 5674 \AA).

Finally, there is a near match between the He~II $\lambda$ 2733 \AA\ line
at the redshift of the z=1.374 C~IV system.  Interestingly, there is
a second He~II line, at $\lambda_0$=2390 \AA\ that would also be a near
match to the observed emission line at 5674 \AA.  However, there
are several pieces of evidence against this identification.  First,
these He~II lines are generally very weak, and rarely seen in
local galaxy spectra (Kinney et al. 1993).  Most importantly, however, 
the relative wavelength
ratios of the helium lines do not match the observed ratios (even accounting
for possible sky line contamination);
since these emission lines originate from the same species, this would be
a fundamental requirement for this match.

{\it (b) The emission line at 5674~\AA.}  In contrast to the above,
there is no bright emission line expected at this wavelength for {\it any} of
the redshifts of the absorbers towards this QSO.  The line has a FWHM of
15~\AA; which corresponds to an effective $(15^2 - 5.2^2)^{\frac{1}{2}} =
14$~\AA\ FWHM once the 5.2~\AA\ instrumental profile is subtracted in
quadrature.  This is equivalent to $\sim 750$ km/s, and could be
caused by either an unresolved multiplet, or from physical broadening
of the line (e.g., from AGN activity). Although there are no obvious
identifications for this emission line associated with the known intervening
absorbers, there are some known fainter lines which may match. 
The first feasible possibility is [NeIV] 2423, 2425~\AA\ at $z
\sim 1.328$, which would be expected $\sim$30~\AA\ ($\sim 1600$ km/s)
blueward of the observed line.  Although usually faint in galaxies,
this transition can be relatively strong in some narrow line radio
galaxies (McCarthy 1993).  However, AGN activity is usually accompanied
by other strong emission lines, such as Mg~II which we had previously
discounted as the identification of the 6505 \AA\ line to the
lack of a doublet feature.  

The second close match is the CII] multiplet 2324--2328~\AA\ at 
$z = 1.446$ which 
would be expected $\sim$17~\AA\ (900 km/s) to the red of the observed line.
The ground state doublet CII] 2324, 2325~\AA\ has an $f$-value ratio 
that is $\sim$ 18:1, therefore the doublet blend is unlikely to account for 
the broadened line, since the emission is so faint.  The excited state
transition CII] $\lambda$2326 is seen in some
AGN (e.g., McCarthy 1993), but again we see no associated Mg~II emission.
Finally, as discussed above, there is a reasonable match with 
He~II $\lambda$2390, but this is also excluded (see previous paragraph).
Therefore, none of these possibilities
presents a very convincing case, suggesting that this emission line
is associated with an object not known in absorption.  Finally, it
is possible that this feature might be due to an
internal reflection in the spectrograph -- however, the fact that it
is coincident with G5, and that it is noticeably more extended than a
PSF makes this an unlikely possibility.  

It is, of course, plausible that the detected emission lines do not
correspond to a known absorber.  Given that we have argued (based on
spatial and kinematic differences) that the two detected emission
lines arise in different objects, an unequivocal identification of
the 6505~\AA\ line as [OII]~$\lambda$3727~\AA\ is not possible.
Lilly et al. (1995) explored the problems associated with redshift
confirmation for single emission line objects in the CFRS and defined
a break index to test for [OII]~$\lambda$3727~\AA, effectively measuring
the presence of a Balmer break.  Although our spectrum is not of
sufficient quality to reliably perform this test, there is a
small rise in flux to the red of the emission line, suggestive
of a possible break that is qualitatively similar to Figure 2
of Lilly et al. (1995).  Finally we note that the other main
emission lines regularly observed in moderate redshift galaxies
can be confidently ruled out - H$\alpha$ $\lambda$6563 would have to
be blueshifted and the [OIII]$\lambda\lambda$4959, 5007 doublet would
be resolved in a clear 1:3 ratio.  If the observed emission line at
6505 \AA\ were  H$\beta$ $\lambda$4861, we would also have observed
[OIII]$\lambda\lambda$ 4959, 5007.

To summarise this subsection, after an exhaustive analysis of
possible emission line matches, we have identified the line at
6505~\AA\ as [OII]~$\lambda$3727~\AA\ emission from G5 at the redshift 
of the $z
= 0.7450$ absorber.  This confirms the galaxy identification of Le
Brun et al.\ (1997) and Bond et al.\ (2001a).  The exact
systemic redshift depends on which wavelength is taken as the central
point of the rotation curve.  Since our S/N is low, this is
difficult to determine accurately, so we adopt $z_{sys} = 0.7450$ for
the rest of the paper.  We find that the
additional emission line seen at 5674~\AA\ is not at the same redshift,
but are unable to find a convincing match with any known absorption
galaxy.  Regardless of the identification,
if we accept this line as real (rather than an artifact) it implies
that the image of G5 is a superposition of a galaxy associated with
the $z = 0.7450$ Mg~II absorber, and a galaxy or AGN at a different
redshift.

\subsection{Internal Kinematics of G5 and the Tully-Fisher Relation}

\begin{figure}
\centerline{\rotatebox{270}{\resizebox{0.85\textwidth}{!}
{\includegraphics{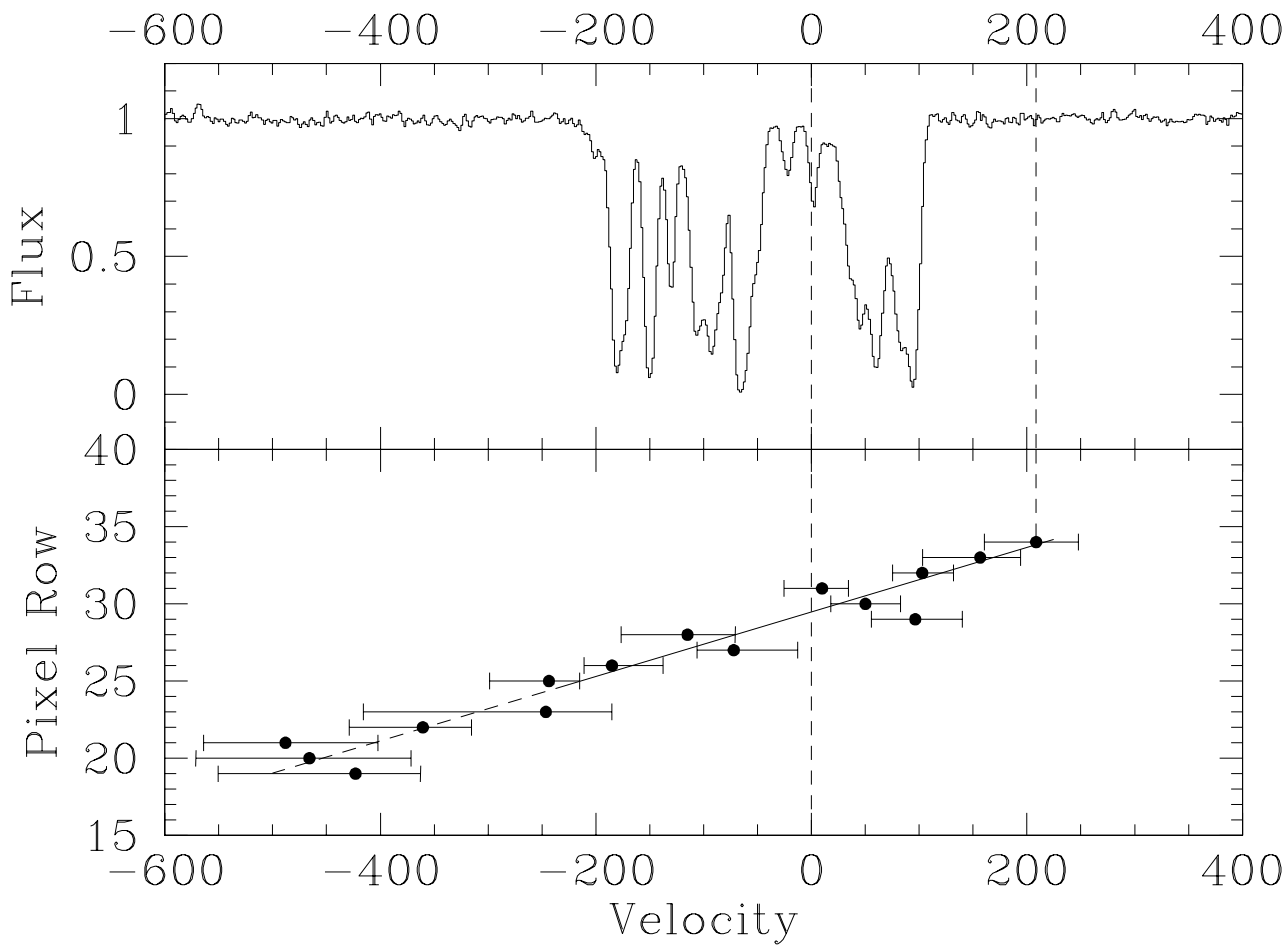}}}}
\caption{Emission and absorption kinematics shown on a common
velocity scale relative to $z=0.7450$.  The emission velocities
were calculated by cross-correlating with an emission line template as
described in the text.  A least squares fit for pixels 25-35 is
shown by the solid diagonal and extrapolated to pixels 19-24
(which are more affected by sky subtraction) in the dashed line. 
The vertical dashed lines indicate zero velocity (corresponding
to the systemic redshift) and $v=208.5$ km/s at the limit of
the detected rotation curve.  Comparison with the UVES data show
that there is no absorption at this velocity in the QSO spectrum.}
\label{em_abs}
\end{figure}

Each line of pixels in the G5 spectrum was cross-correlated with a template of
the [OII]~$\lambda$3727~\AA\ doublet, which was constructed by 
co-adding two arc
lines with the separation corresponding to 3726.032, 3728.815 \AA\ in a 1
to 1.3 ratio at a redshift of $z = 0.7450$ .  
Error bars were estimated from a simulation in which 
random Gaussian noise was added to each line of the spectrum 100 times, 
and the cross-correlation with the template was performed on each noisy 
spectrum.  The resulting velocity points and errors are shown in
Figure~\ref{em_abs} and exhibit a linear velocity gradient (a similar
test on the unidentified emission line at 5674 \AA\ exhibits no
velocity structure).  While the velocities in the
upper region of the emission line (above pixel 25) are reliable,
the lower region lies in a part of the spectrum which originally
had strong sky emission, and the observed [OII] line and its 
velocity are therefore suspect.  In particular, the line profiles for the 
lower portion of the emission line are skewed towards shorter wavelengths,
therefore giving a cross-correlation velocity which is
likely blue-shifted from the true velocity (compare the
lower portion of Figure 2d to the lower panel of Figure \ref{em_abs}).
We perform a least squares fit to the velocity points, shown
as a diagonal line in Figure \ref{em_abs}.  The errors on the 
least squares fit yield a formal error of $\pm$32 km/s on the rotation
velocity.  However, in practice we are restricted to quoting a
lower limit to $v_{rot}$ by the low S/N at the lower end of the curve,
possible contamination by sky line residuals
and by the fact that we may well not be seeing out to the flat part
of the rotation curve.  We also note that convolution of the emission
line with the seeing profile will also serve to under-estimate $v_{rot}$
by blending signal from stronger emission
towards the inner part of the disk at lower velocities.
The straightforward interpretation of these data is that G5 is a large
disk galaxy with $v_{rot} \gtrsim 210$ km/s (determined from the
upper portion of the emission line) and possibly as large as
350 km/s if the lower extension of the [OII] line is real.

As a second estimate of the rotation velocity,
we calculate $v_{rot}$ directly from the Tully-Fisher relation (e.g.,
Tully \& Fisher 1977), based on the absolute magnitude of the galaxy,
as derived from our re-appraisal of the HST data.  In order to compare
with published Tully-Fisher relations locally and at intermediate
redshift, we convert the $M_B(AB)$ calculated for our preferred
cosmology to those adopted by Vogt et al.\ (1997) and Kannappan, Fabricant
\& Franx (2002), i.e.,  $q_0$ = 0.05
and $H_0$=75 km/s/Mpc.  We further correct this value of $M_B(AB) =
-19.04$ by +0.14 mags to convert to the Vega scale and by $-0.67$ mags
following the prescription of Tully \& Fouqu\'e (1985) to correct for
internal extinction for an edge-on galaxy.  The resulting magnitude
after these corrections is $M_B = -19.57$.

We derive the expected velocity for a galaxy of this magnitude 
using the Tully-Fisher relation from Kannappan, Fabricant
\& Franx (2002) (KFF).  We calculate the maximum rotation velocity
by taking the unweighted inverse fit
from Table 1 of KFF $M_B = -19.83 - 10.09 \times (logW_{50} - 2.5)$,
where $W_{50}$ is the 50\% HI 21-cm linewidth.  The linewidth is related 
to the rotation velocity $v_{max}$ by
$W_{50} = 19 + 0.90 \times 2 v_{max}$ with scatter 25 km/s (KFF appendix
B).  Using our $M_B = -19.57$ value for G5, we obtain 
$W_{50} = 298 \pm 40 $ km/s, and $v_{max} = 155 \pm 40 $ km/s. 
In order to reproduce the observed  (minimum) rotation velocity 
of $v_{rot} = 210$ 
km/s, the same relationship from KFF yields an absolute $B$-band
magnitude of $-20.83$, i.e., 1.26 magnitudes brighter than the value
that we measure (once the edge-on dust correction of Tully \& Fouqu\'e 
1985 is made). If the lower extension to the rotation curve is real,
then the rotation velocity could be as large as 350 km/s, which would
imply $M_B = -22.98$, i.e., require an additional 3.4 magnitudes
of extinction after the edge-on correction in order to match the Tully-Fisher
relation.
Therefore, there is a significant discrepancy between our measured luminosity
and rotation curve with the local Tully-Fisher relation, which would
be further exacerbated by including the lower extension of the emission
line.  

There is still considerable debate over the possible evolution of the
Tully-Fisher relation at $z>0$.  Vogt et al.\ (1996, 1997) found that
out to $z \sim 1$ there is relatively little evolution ($< 0.4$
mag brightening) for large disk galaxies to z$\sim$1, consistent with
the HST WFPC2 morphological study of Lilly et al., (1998).  Other
groups, however, have found significant luminosity evolution in small,
blue galaxies (e.g., Phillips et al.\ 1997, Rix et al.\ 1997, Simard
\& Pritchet 1998, Mall\'en-Ornelas et al.\ 1999), and even in large
spirals (e.g., Rigopoulou et al.\ 2002).  Ziegler et al.\ (2002) have
suggested that the extent of luminosity evolution depends on rotation
velocity with significant corrections required for 
galaxies with $v_{rot} < 150$
km/s.  Despite this still murky picture, we can nonetheless make a
clear statement about the case of G5. All suggestions of evolution in
the Tully-Fisher relation advocate brightening (for a given $v_{rot}$)
at higher redshifts; if present this would further increase the
discrepancy between our measured luminosity, and that predicted by the
Tully-Fisher relation for our measured $v_{rot}$.

\subsection{Absorption Line Profiles}

In order to determine column densities for the metal species,
we have used the Voigt profile fitting program VPFIT\footnote{Available 
at http://www.ast.cam.ac.uk/\~{}rfc/vpfit.html } to determine column
densities for the individual absorption components.   We have adopted 
the common strategy of tying Doppler widths ($b$-values) and redshifts 
of components  for different species and obtain good simultaneous fits 
for Mg~I, Mg~II and Fe~II indicating that these species occupy the same 
velocity space.   Fit parameters are given in
Table \ref{vp_fits} and we note that since several of the
Mg~II components are mildly saturated, we quote the total
N(Mg~II) as a lower limit.
In principle, HST spectra are required in order to determine the 
N(H~I) of this system.  However, in practice, the Lyman limit of 
the higher redshift DLA prevents us from measuring the Ly$\alpha$
absorption in the lower redshift system.   
However, as discussed by Bond et al.\
(2001a), although the Mg~II $\lambda$2796 and Fe~II $\lambda$2600 
equivalent widths (EW) in this absorber give a 50\% probability that
this is a DLA (Rao \& Turnshek 2000), the absence of
extended Mg~II saturation would be unusual for a \textit{bona fide}
damped system.  Due to the lack of N(H~I) constraint, we are unable to 
determine abundances for this absorber.  However, we return to the issue 
of relative column densities below.  

\begin{figure}
\plotone{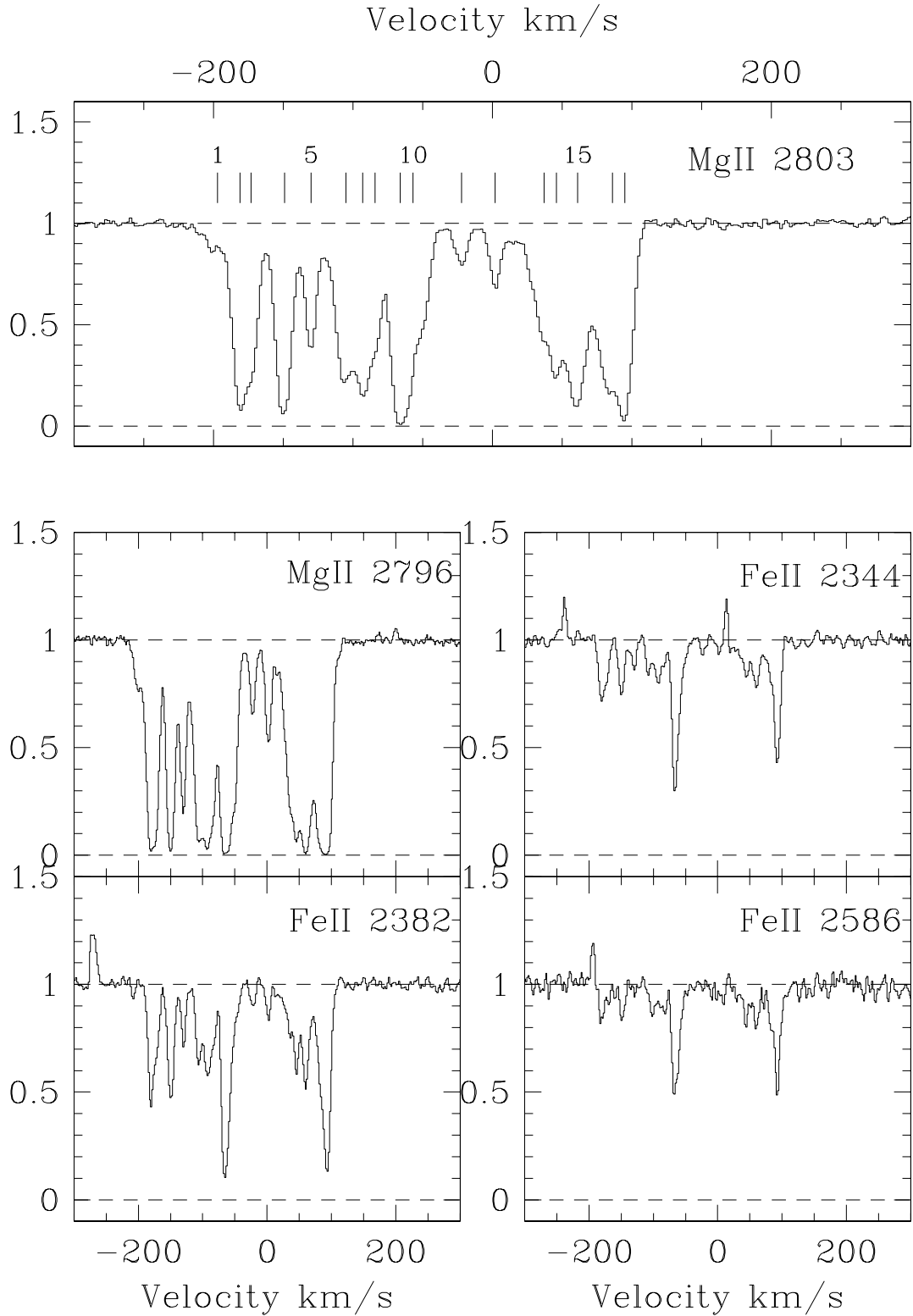}
\caption{FeII and MgII lines covered by UVES data plotted on a common
velocity scale relative to $z_{abs}=0.7450$. \label{uves_data}}
\end{figure}

The absorption profiles of this system are extended over approximately
300 km/s.  We adopt the systemic absorption redshift to be the
same as that determined for the [OII] $\lambda$3727 \AA\ emission line,
i.e., $z=0.7450$.  Figure \ref{uves_data} shows the Mg~II and Fe~II 
lines covered by the UVES data.  
The slightly higher resolution of the UVES data
compared with the HIRES spectra of Bond et al.\ (2001a) reveals even 
more striking symmetries in the absorption features.  However,
the clearest examples are now not the pairs that are separated
by $\sim 30$ km/s, but those with much larger velocity differences
e.g.,  the strongest components at $\sim -50$ km/s and 
$\sim +100$ km/s.

\begin{figure}
\centerline{\rotatebox{0}{\resizebox{0.85\textwidth}{!}
{\includegraphics{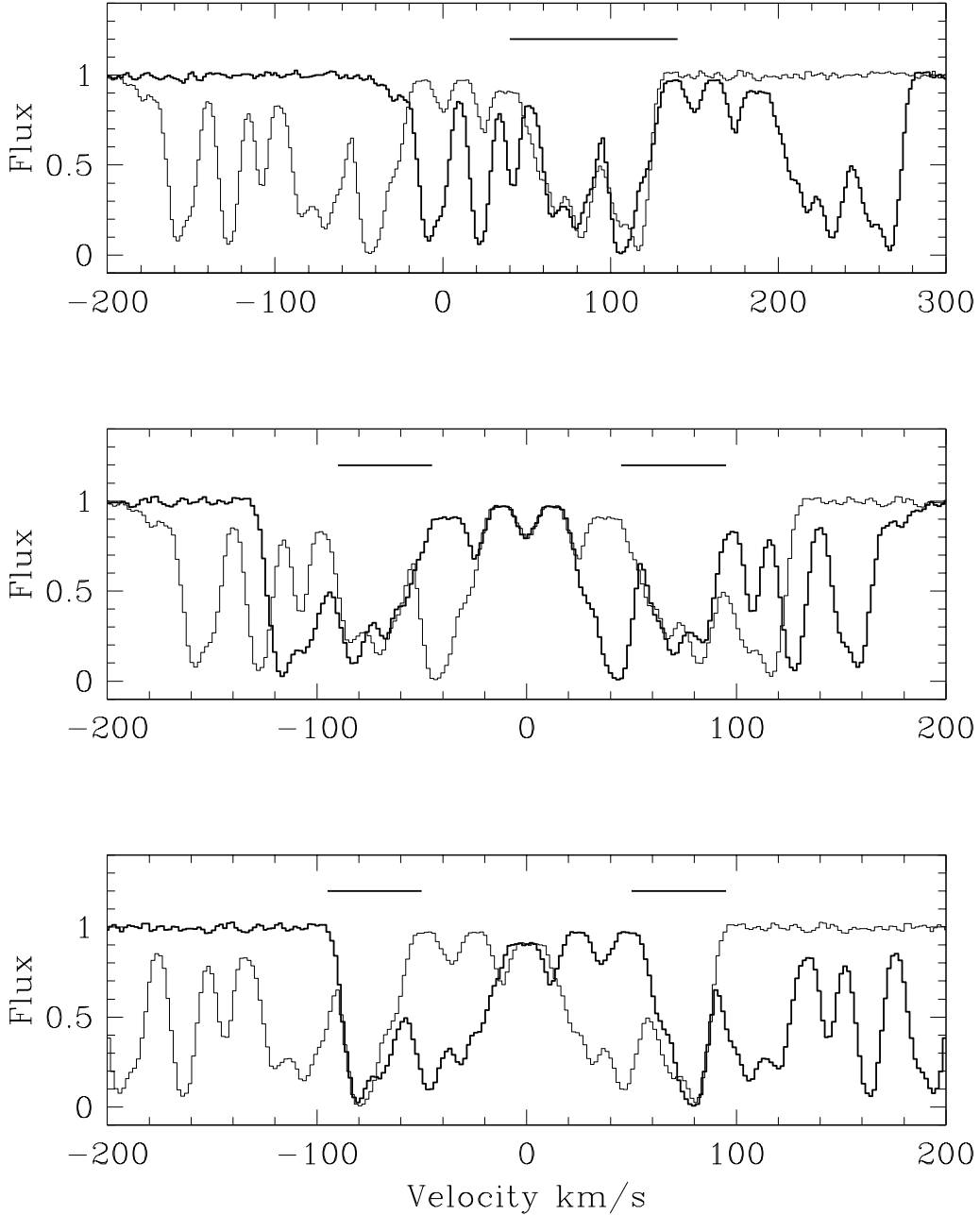}}}}
\caption{Figure to illustrate the absorption profile symmetries in
the Mg~II system towards Q1331+17.  In all panels, horizonal bars are
a guide to the symmetries illustrated.  Top panel:  the Mg~II 
$\lambda$2803 line is shown on a velocity scale relative to
 $z_{abs}=0.74487$ (thin solid line) and redshifted by 150 km/s 
(thick solid line) to illustrate the `doublet symmetry'.
Middle panel: the Mg~II $\lambda$2803 line is shown on a velocity scale 
relative to  $z_{abs}=0.74487$ (thin solid line) and with the 
velocity scale flipped (thick solid line) to illustrate the mirror 
symmetry of components at $\sim \pm 80$ km/s.  Bottom panel: the Mg~II 
$\lambda$2803 line is shown on a velocity scale 
relative to  $z_{abs}=0.74508$ (thin solid line) and with the 
velocity scale flipped (thick solid line) to illustrate the mirror 
symmetry of components at $\sim \pm 80$ km/s. }
\label{symm}
\end{figure}

In Figure \ref{symm} we illustrate qualitatively some of these symmetries with
the Mg~II $\lambda$2803 transition.
The top panel shows the spectrum on both a velocity scale relative
to $z=0.74487$ and
also shifted by 150 km/s to the blue demonstrating the keen similarities 
in terms of velocities between components 6 -- 10 and 13 -- 17, as
marked by the horizontal bar.  The middle panel shows the spectrum 
both relative to $z=0.74487$ and also flipped in velocity space to
illustrate the `mirror' symmetries.  
Notice how the pairs of lines at $\sim \pm 80$ km/s (components 6, 7 \&
8 compared with 13, 14 \& 15) overlap, including the broadening towards 
zero velocity.  The bottom panel illustrates another example of such
a mirror symmetry, this time between components 9 and 10 compared with
16 and 17.   With a single line of sight, we are not able to further
constrain the possible geometry of this system (c.f. Rauch et al. 2002), 
but since patterns of 
this sort may be expected if the line of 
sight intersects an expanding shell, we speculate that the components
6 -- 17 may arise as a consequence of bubbles (see also Bond et al.\
2001a).

We can, however, assess whether the observed symmetries could plausibly arise
by chance due to a random placement of lines by using the Two Point
Correlation Function (TPCF) and Column Density Distribution Function
(CDDF) for Mg~II systems from Churchill, Vogt \& Charlton (2002).  The TPCF
quantifies the probability of finding 2 absorption components with a
separation $\Delta$V and is parameterised by a double Gaussian of the form

\begin{equation}
P(\Delta V) = A (2 e^{-(\frac{\Delta V^2}{2 \sigma_1^2})} + e^{-(\frac{\Delta V^2}{2 \sigma_2^2})} )
\end{equation}

\begin{figure}
\centerline{\rotatebox{0}{\resizebox{0.85\textwidth}{!}
{\includegraphics{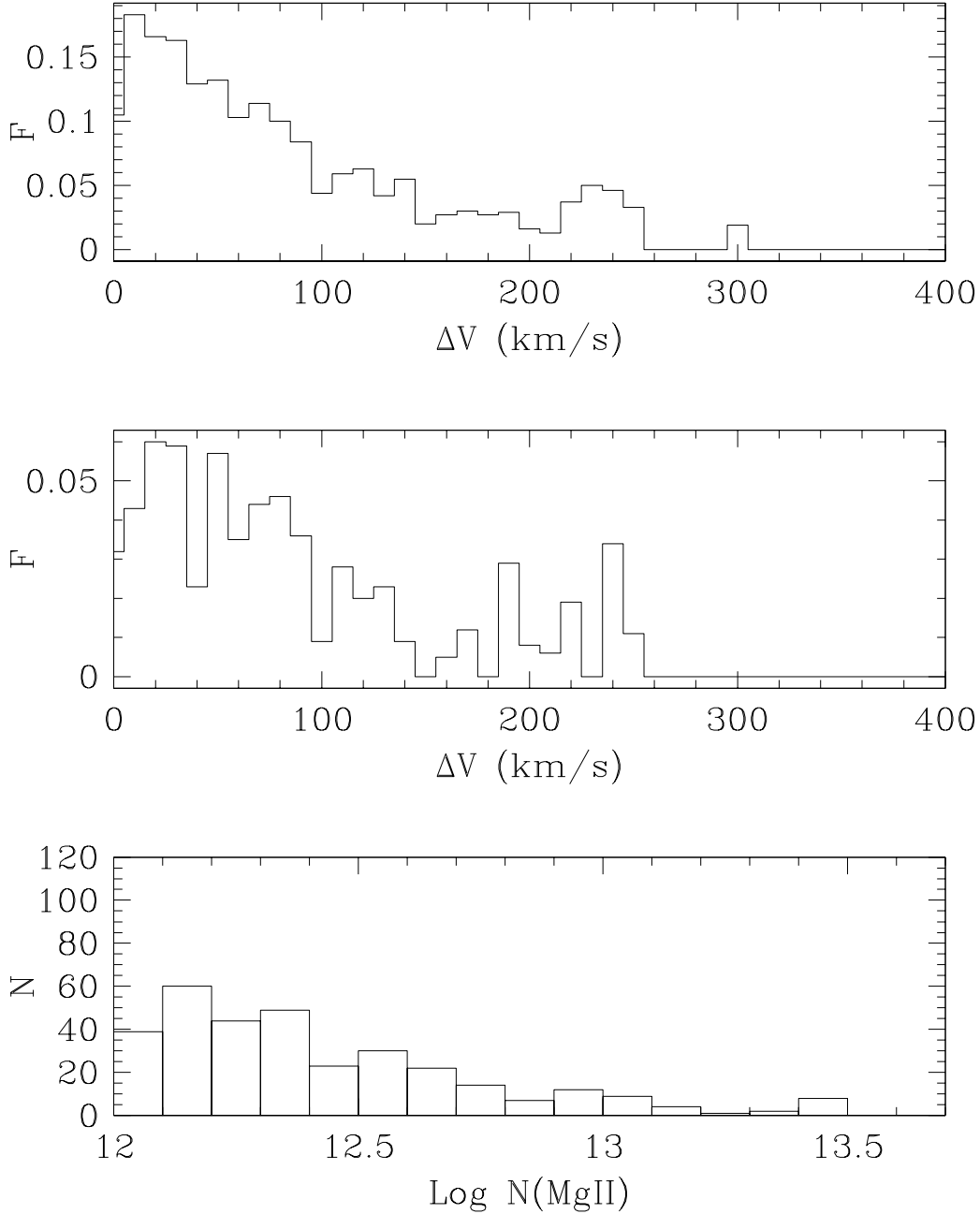}}}}
\caption{Results from 10000 realisations of two Mg~II pairs drawn from
the velocity and column distributions parameterised by the TPCF and CDDF
of Churchill, Vogt \& Charlton (2002).  Top panel:  Fraction of pairs
with matching velocity separations (within a tolerance of 10 km/s)
as a function of $\Delta$V.  Middle Panel:  Fraction of pairs with
matching $\Delta$V \textit{and} matching column densities (within a 
tolerance of 0.3 dex) as a function of $\Delta$V.  Bottom panel:
Absolute number (from the 10000 realisations) of matched lines with
a given column density.  These simulations show that although we expect
matches with velocity separations of a few 10 km/s~ $\sim$ 5\% of the
time, very few will have the moderately high column densities observed
towards Q1331+17.\label{test_symmetry}}
\end{figure}

From their large sample of Mg~II systems, Churchill, Vogt \& Charlton (2002)
have determined $\sigma_1$ = 54 km/s and $\sigma_2$ = 166 km/s.
The CDDF describes the number of components with a given column density
and is a power law of the form

\begin{equation}
f(N) \propto N^{\beta}
\end{equation}

with $\beta=-1.59$ for Mg~II components.  By combining the TPCF and the 
CDDF we can assess the probability that two
pairs of components in a given system will have the same $\Delta$V and matching
column densities, within some tolerance (we make no distinction between
`shifted' symmetries and `mirror' symmetries in these calculations).  We
produce 10000 realizations of 2 pairs of Mg~II lines and apply a relatively
relaxed tolerance that $\Delta$V $<$ 10 km/s and $\Delta$logN $<$ 0.3 dex
\footnote{The symmetries we observe actually match somewhat better than this,
see Table 2.  However, the errors on the individual component column densities
can be several tenths of a dex, hence our conservative tolerances.}.  
The results are shown in Figure \ref{test_symmetry}.  In the top panel, 
we plot the fraction of pairs that have the same $\Delta$V, as a function 
of $\Delta$V.  Since small values of $\Delta$V are more probable,
this fraction is larger for smaller velocity separations.  
In the middle panel, we plot the fraction of pairs which have matching 
$\Delta$V \textit{and} logN, again as a function of $\Delta$V.  
The typical $\Delta$V between components of the symmetric pairs 
observed in Q1331+17 is $<$ 15
km/s, so the results of these simulations (Figure \ref{test_symmetry}, 
middle panel) would
imply that in a few percent of cases we may expect to find \textit{one set}
of matching pairs.
However, this is a strong function of column density; since the CDDF is a
power law with negative slope, high column density components are
intrinsically rarer and therefore matches that include a high column
density component will also be rare.  Indeed, several of the symmetric
components in Q1331+17 have relatively large column densities, log
N(MgII)$>$12.7.  In the bottom panel of Figure \ref{test_symmetry} we
have plotted a histogram showing the absolute number of matches (for 
10000 realizations) as a function of column density.  Clearly,
matched components with moderately large N(Mg~II) should be extremely rare.
Moreover, these simulations only consider matches between two pairs
of lines whereas in the spectrum of Q1331+17, we see multiple pairings, 
sometimes with more than two components.  While the probability of 
chance pairing of two lines is, as shown by
our simulations, already small, the pairing of whole line complexes is
even more unlikely; its probability can be approximated by the
probability of the pairing of N individual line pairs, and therefore
decreases quadratically compared to the already small probability of
chance pairings of individual lines. Therefore, the likelihood
of observing symmetries such as those seen in this Mg~II system
are vanishingly small ($<$0.02\%).

\section{Discussion}

\subsection{Galaxy Association and Kinematics}

Steidel et al.\ (2002) have recently studied the absorption and
emission kinematics of 5 Mg~II absorbers associated with inclined disks.
They find that 4/5 of these systems are consistent with absorbing gas
that is co-rotating with the main galactic disk (although 
the profiles are not consistent with pure disk rotation).
In the Mg~II absorber towards Q1331+17, we find that the absorption
components fall approximately central to the systemic redshift 
of the emission line, but
with a large spread ($\sim$ 100 km/s) to either side (Figure \ref{em_abs}).
Although Steidel et al.\ (2002) find one example of absorption
gas located at the centre of the rotation curve, the velocity
spread of the absorption components is very small.  Whilst we do not
detect the turnover of the rotation curve (and even with the conservative
assumption that the curve flattens beyond the limit of the detected
rotation) it is clear 
from Figure \ref{em_abs} that there is \textit{no absorption at the
extrapolated disk rotation velocity determined for G5}.  In fact,
Figure \ref{em_abs} implies that if all of the absorbing gas were
associated with a disk, then a significant amount would have to be at
counter-rotating velocities.  Interestingly, Lanzetta et al.\ (1997)
have reported a similar incidence of counter-rotating gas in a 
$z_{abs} = 0.1638$ Lyman limit system associated with a 0.4$L_{\star}$ 
spiral inclined at $\sim 50^{\circ}$.  

Although several examples of counter-rotation between stars and
gas have been 
studied in the local universe (e.g., Galletta 1987;  
Bertola, Buson \& Zeilinger 1992; Ciri, 
Bettoni \& Galletta 1995; Jore et al.\ 1996; Rubin, Graham \& Kenney 1992), 
kinematic de-coupling remains a relatively rare phenomenon among spirals
(Kannappan \& Fabricant 2002).  
Given the even higher scarcity of outer disk vs. inner disk gas decoupling
(e.g., Braun, Walterbos, Kennicutt 1992), it would be extremely surprising
that two (Q1331+17 and Q0850+44, Lanzetta et al.\ 1997) out of
8 (i.e., including the 5 additional systems from Steidel et al.\
2002 and the system of Barcons et al. 1995) galaxy absorbers need to 
appeal to this explanation.  A more
likely explanation in the case of the $z=0.7450$ Mg~II absorber
towards Q1331+17 is that internal non-disk kinematics play
a dominant role, consistent with the scenario of bubbles or
winds.  Alternatively, the absorption could be associated with
a companion to G5 (i.e., at the same redshift) which would make
its relative velocity entirely independent of G5's rotation curve.
Zaritsky et al.\ (1997) have shown that the velocity distribution
between local spirals and their satellites extends to several hundred
km/s, although most are within 200 km/s (but mostly along the disk's
minor axis).   Lanzetta et al.\ (1997) also 
raise the possibility
that a faint nearby galaxy could be causing the
absorption in the case of Q0850+44.  Their argument is further supported
by the absorber's very low metallicity ($ < \frac{1}{25} Z_{\odot}$),
which is much lower than would normally be expected for a 
0.4$L_{\star}$ based on the metallicity-luminosity relation (Kobulnicky
\& Zaritsky 1999).  Unfortunately, due to the impossibility of an
N(H~I) measurement for the absorber in Q1331+17, we can not make
a similar comparison for the Mg~II system studied here.

Although G5 is fainter than the typical Mg~II galaxy (Steidel,
Persson \& Dickinson 1994), its luminosity and impact parameter are
consistent with the gas cross section relation derived by Steidel
(1995), $R = 35 h_{100}^{-1} kpc (L/L_{\star})^{0.2}$.  Whilst the
correlation between Mg~II equivalent width and luminosity is weak
(Churchill et al.\ 2000b), it is nonetheless striking that the
absorber towards Q1331+17 is one of the strongest compared with the
sample of Churchill et al.\ (2000a), yet it is at the faint end of the
luminosity distribution.  This could be a further symptom of
superbubbles which would increase the density of local gas and
consequently increase the N(Mg~II) for a given metallicity.  This
possibility could be tested by identifying the absorbing galaxies
associated with other potential superbubble Mg~II candidates presented
in Bond et al.\ (2001b).

A re-assessment of the archival WFPC2 imaging has led to a significant
downward revision of G5's luminosity to $L \sim 0.3 L_{\star}$.
Compared with the rotation velocity of $v_{rot} \gtrsim 210$ km/s, this
is inconsistent with the Tully-Fisher relation, both locally and
at higher redshifts.  We can speculate upon various possibilities
that would reconcile this inconsistency.    The identification
of a second emission line, indicates that a second galaxy is present
coincident (at least to some degree) with G5.  Although the S/N
of G5 in the WFPC2 image is low, there is evidence for a faint
`kink', see Figure \ref{fors_mask}b.  This will undoubtedly
affect the luminosity of G5, although it would lead to an
even fainter absolute magnitude. Therefore, appealing to
luminosity evolution (\S 3.2) or blending with another coincident
object only exacerbate the Tully-Fisher discrepancy.  It is
possible that our internal extinction correction may be 
inaccurate and large amounts of dust are known to exist
both in and above the plane of disk galaxies (e.g., Howk
\& Savage 1999).  However, the total extinction correction required
to reconcile the luminosity with the Tully-Fisher relation is 
approximately 3 times the maximum given by Tully \& Fouqu\'e (1985),
more if the true rotation velocity is significantly larger than
our current lower limit of 210 km/s.  We also note that 
none of the 5 galaxies studied by Steidel et al.\ (2002) exhibit such 
a discrepancy with the Tully-Fisher relation.
It is also possible that G5 truly lies a long
way from the classical Tully-Fisher relation.  For example, 
there is some evidence that low surface brightness (LSB) 
galaxies deviate from the classical Tully-Fisher relation in the sense
that some are apparently
under-luminous for the rotation velocities, e.g., O'Neil et al.\ (2001).
However, more recent work suggests that the majority of discrepant
points may suffer from H~I contamination from nearby galaxies
(Chung et al.\ 2002). Our interpretation
of superbubbles (see discussion in the next section)
signals active physical processes such as 
intense star formation or even merging which may impact significantly
on the relation between luminosity and rotation velocity.
Intense bursts of star formation will make a galaxy
even brighter for a given $v_{rot}$, again exacerbating the
Tully-Fisher discrepancy we observe, although the global effects
on the Tully-Fisher relation of merging or starburst galaxies are unknown.  
Unfortunately, these data are not of sufficient quality to
determine a star formation rate with a meaningful accuracy that
would allow us to explore these possibilities in any quantative detail.
Finally, it is important to stress again that G5 is a faint galaxy
with weak emission lines, so measurement errors are relatively
large.    Similarly, the Tully-Fisher exhibits
a significant scatter.  As calculated above, the absolute $B$-band
magnitude error would have to be under-estimated by a factor of $\sim$
6, but the large error on the rotation velocity places it within 
a 1 $\sigma$ scatter from the Tully-Fisher relation of 
Kannappan, Fabricant \& Franx (2002).

\subsection{On The Presence of Superbubbles}

The pair-wise absorption pattern noted by Bond et al.\ (2001a) is
even more striking in the UVES data presented here (see Figures 
\ref{uves_data} and \ref{symm})
and is indeed reminiscent of the expected signature of expanding gas
shells.  Whereas Bond et al.\ focussed on line pair separations
of $\sim$ 30 km/s, the UVES data show even more striking symmetries
between components with $\Delta v \sim 150$ km/s, such as between
components 6, 7 \& 8 with components 13, 14 \& 15.
It is interesting to note that several local cases
of absorption in QSOs behind readily identifiable galaxies
may be associated with disturbed gas, including bubbles, shells and mergers
(e.g., Norman et al.\ 1996; Womble, Junkkarinen \& Burbidge 1992 and 
references therein).  Bond et al. (2001a) used CLOUDY models to
infer the energetics of the starburst that may be driving the bubble
and the implied metallicity of the gas.  These models place limits on
the energy injected into the bubble and its age, assuming a metallicity
and an empirically motivated range of column densities and shell 
thicknesses.   With a $\Delta v$ larger by a factor
of 5 than those studied by Bond et al. (2001a), these models would be 
shifted to energies some 300 times larger for a given age
(see Bond et al. 2001a for
the formalism of these calculations).  The combined effect of metallicity
and starburst luminosity governs the shell thickness and observed Mg~II
column density.  Extrapolation of the models in Bond et al. (2001a)
implies total luminosities of between 1 -- 10 $\times 10^{38}$ ergs 
cm$^{-3}$ corresponding to an OB association of some 500 -- 5000 stars,
and a metallicity of 1/10 solar.  Given that this upper limit is similar 
to the largest OB associations seen at $z=0$ (e.g. McKee \& Williams
1997), this implies that lower metallicities are unlikely.  Higher
metallicities could be reproduced by proportionately lower energies.

\begin{figure}
\centerline{\rotatebox{0}{\resizebox{0.85\textwidth}{!}
{\includegraphics{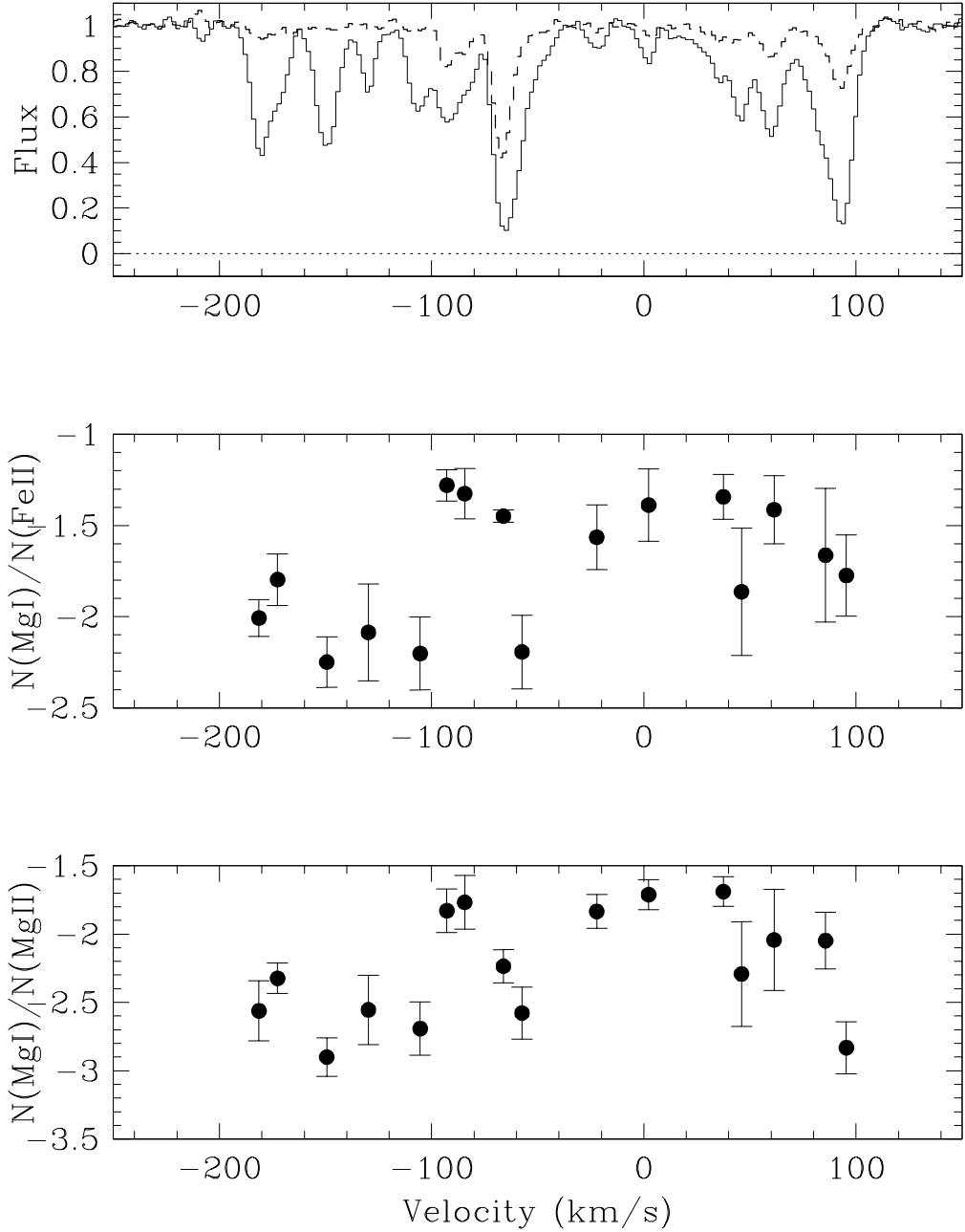}}}}
\caption{Top panel: FeII $\lambda$2382 (solid line) and MgI (dashed line)
plotted on a common velocity scale.  Middle panel: N(Mg~I)/N(Fe~II)
on a component by component basis, taken from the fits in Table 
\ref{vp_fits}.  Bottom panel: N(Mg~I)/N(Mg~II) on a component by 
component basis, taken from the fits in Table \ref{vp_fits}. Note
that since many of the Mg~II components are mildly saturated, these
ratios are upper limits.  The error bars are those produced by VPFIT
for individual components and are largest when components are
severely blended.}
\label{density}
\end{figure}

\begin{figure}
\centerline{\rotatebox{270}{\resizebox{0.85\textwidth}{!}
{\includegraphics{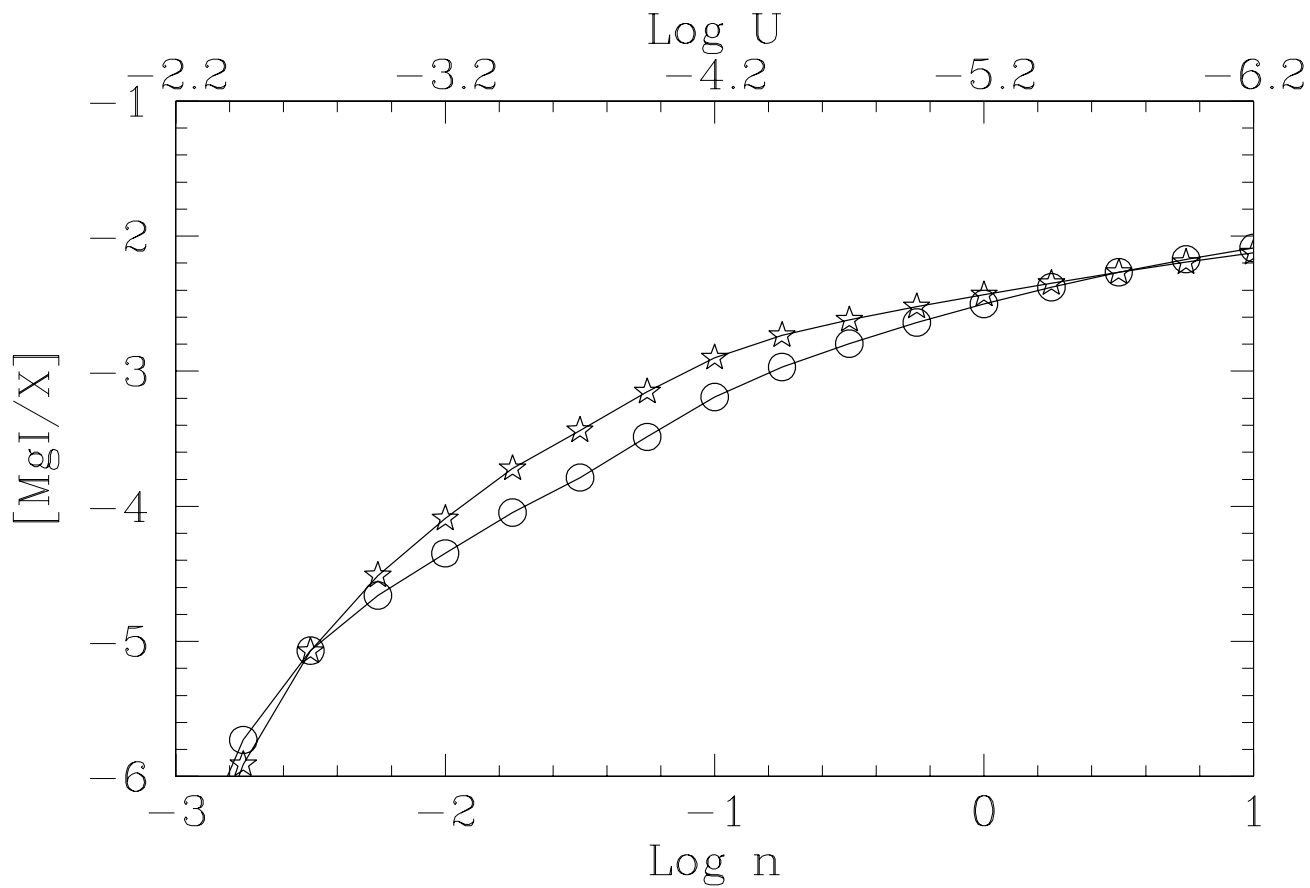}}}}
\caption{The ratio of N(Mg~I)/N(Mg~II) (stars) and N(Mg~I)/N(Fe~II) (circles)
as a  function of $n$, the total hydrogen density (cm$^{-3}$) for log 
N(H~I)=19.   These values were determined
using a CLOUDY94 model with a plane-parallel gas slab with metallicity
$Z=-1.5$, a power law ionizing spectrum with $\alpha=-1$, normalised to 
$J_{21}$=0.1 at 1 Ryd.}
\label{cloudy}
\end{figure}

The ratios of Mg~I/Fe~II and Mg~I/Mg~II contain information
pertaining to the ionization state (and hence, density) of the
gas and may provide clues to the nature of the absorption.
In Figure \ref{density} we plot the component by component Mg~I/Fe~II
and Mg~I/Mg~II ratios (along with the velocity-scaled absorption
profiles to guide the eye).    Ideally,
we would like to compare N(Mg~I)/N(Mg~II) and indeed there is
an interesting trend that the components that exhibit the symmetries
described above (i.e., those components we have tentatively identified as 
arising in bubble shells) generally have higher N(Mg~I)/N(Mg~II).  
Although Churchill, Vogt \& Charlton (2002) have recently shown that
N(Mg~I)/N(Mg~II) correlates with N(Mg~II), the distribution
of N(Mg~I)/N(Mg~II) in Figure \ref{density} is independent of this
relation.  However, since many of the Mg~II components are partially
saturated, these 
ratios are often upper limits.   The ratio of N(Mg~I)/N(Fe~II) is 
unaffected (in this case) by saturation, although we have no \textit{a priori} 
knowledge that the Mg/Fe ratio is identical in every component 
(e.g., Ledoux, Bergeron \& Petitjean 2002; Ellison 2000) there
is recent evidence that the abundance ratios, at least in DLAs,
show remarkable uniformity (Prochaska 2003; Churchill et al. 2003).  
Nonetheless, 
the N(Mg~I)/N(Fe~II) is almost universally higher in the components 
proposed to be bubbles, indicative of large densities.  Column
densities of highly ionized species such as C~IV would help to
constrain the ionization parameter, however the only existing
HST spectra of Q1331+17 are of very poor quality (Bechtold,
private communication).  Rauch et al.\ (2002)
find similarly high ratios of N(Mg~I)/N(Mg~II) and N(Mg~I)/N(Fe~II)
in an absorber at $z_{abs}=0.5656$ which they have interpreted as
due to superbubbles.  In addition, similar column density ratios of these
species have been found in low redshift
Mg~II absorbers known to be undergoing strong galaxy-galaxy interactions
(Norman et al.\ 1996).  In Figure \ref{cloudy} we present  
representative CLOUDY94 (Ferland et al.\ 1998) models of  N(Mg~I)/N(Mg~II)
and N(Mg~I)/N(Fe~II) to illustrate that
large values of this ratio are indicative of high densities.
However, modelling the precise density of absorbers depends on the N(H~I) 
which is 
unknown for this system.   Moreover, given that the bubbles 
may consist of shocked gas that is not in thermal equilibrium and 
that the effective ionizing radiation spectrum is unknown, precise 
modelling of relative abundances is outside the scope of this paper.

We conclude that, although the nature of this absorber can never
be unequivocally determined solely from its absorption properties,
the double-peaked profile of the redward components, their striking
symmetries and high N(Mg~I)/N(Mg~II) and N(Mg~I)/N(Fe~II) ratios 
and large Mg~II equivalent width support the interpretation of 
superbubbles put forward by Bond et al.\ (2001a), but the larger
velocity separations investigated here imply larger energies and
enriched gas.  

\section{Summary}

We have presented complementary quasar absorption line spectra
and spatially resolved emission line data for a candidate Mg~II absorbing
galaxy at $z \sim 0.745$.  Our main conclusions are:

\begin{enumerate}

\item We have obtained a spectroscopic redshift for galaxy G5 based
on the detection of the [OII] $\lambda$3727 line.  The galaxy,
located at 3.86 arcsec (28.3 $h^{-1}$ kpc) from QSO Q1331+17, is
at a redshift $z_{sys}=0.7450$.  There is a second
unidentified emission line which does not match any expected line
at $z_{sys}=0.7450$ or for any other of the known QSO absorption
systems.  In combination with its distinct spatial profile, this
indicates a second object, at a different redshift is (at least 
partially) coincident with G5.

\item A re-analysis of archival
HST data leads us to determine an absolute luminosity 
$L \sim 0.3 L_{\star}$ for this galaxy.  If G5 is an edge-on
disk, our detection of [OII] $\lambda$3727 indicates a rotation
velocity of $\gtrsim$210 km/s.  This is inconsistent
with the established Tully-Fisher relation both locally and
at intermediate redshifts.

\item  The QSO absorption is aligned with the center of the emission
line, i.e., at $v=0$.  There is no absorbing gas at the extension of
the rotation curve, but there is a significant amount at negative
velocities.  We suggest that this is most likely explained due
to internal motions, such as winds or bubbles which may be associated
either with non-disk component of G5 or an as yet unidentified companion.

\item  The double-peaked absorption profiles, striking symmetries
and large N(MgI)/N(MgII) and N(MgI)/N(FeII) ratios are indicative
of expanding superbubbles.  This is consistent with observations
at low redshift which find that extended absorption halos are
often associated with disrupted gas, for example in shells, mergers or
interactions. Superbubbles may also explain why the absorber
towards Q1331+17 has a relatively large Mg~II equivalent width
relative to the luminosity of its absorbing galaxy.  The OB
association driving the bubble may have a total energy of 1 -- 10
$\times 10^{38}$ ergs cm $^{-3}$ and contain as many stars as the
largest seen at the present day.  This indicates that the gas
has been enriched to at least 1/10 solar metallicity.

\end{enumerate}

\acknowledgments

SLE is grateful to the IAS and Princeton University for hosting a visit,
during which part of this work was carried out.  MJS acknowledges
support from Proyecto Fondecyt de Incentivo a la Cooperaci\'on Internacional
No. 7000529 for visits to Chile.  
GMO is partially supported by Fundaci\'on Andes and by Proyecto
Fondecyt Regular 2000 No. 1000529.   The authors thank 
Nick Bond, Joop Schaye, Chris Howk, Howard Yee and Sebasti\'an L\'opez 
for valuable input and 
Chris Churchill, Pat Hall, David Bowen and Todd Tripp 
for stimulating discussions.

%
%

\begin{table*}
\begin{center}
\caption{Possible emission line identifications in the spectrum of G5
based on the wavelength correspondence with absorption line systems in the QSO
line of sight.  Tabulated are the type of absorption system, its redshift, 
the corresponding rest-frame
wavelength range of the FORS2 spectrum, and the rest-frame wavelength
and possible ID for the two observed emission lines at 6505~\AA\ and
5674~\AA.   For the rest-frame wavelength range covered the only
important lines visible are [OII]~3727~\AA\ for the $z = 0.745$ system,
and Mg~II 2796,2803~\AA\ for the systems at $z = 1.328, 1.374, 1.446, 1.776$.
Credible matches are given in bold text (see text for discussion).}
\vspace*{5mm}
\begin{tabular}{|c|c|c|cc|cc|}\hline \hline
Absorption &z &$\lambda$ range&\multicolumn{2}{c|}{6505~\AA\ line} & \multicolumn{2}{c|}{5674~\AA\ line}\\
System       &  & (\AA)         & $\lambda_{rest}$ (\AA)             &Possible matches & $\lambda_{rest}$ (\AA)             &Possible matches  \\
\hline
Mg~II&  0.745 & 3153--4655  & 3729 &	{\bf [OII] $\lambda$3727}&	3253&\\
Mg~II&  1.328  & 2362--3487  & 2794 &	MgII $\lambda\lambda$2796, 2803	&	2437&  [NeIV] $\lambda\lambda$2423, 2425\\
C~IV &  1.374  & 2317--3421  & 2740 &	HeII $\lambda$2733	&	2390& HeII $\lambda$2385 \\
C~IV &  1.446  & 2248--3319  & 2659 &			&	2320&  CII] $\lambda$2324--2328\\
DLA  &  1.776  & 1981--2925  & 2343 &			&	2044&  \\
\hline\hline
\end{tabular}
\label{line_ids}
\end{center}
\end{table*}

\begin{table*}
\begin{center}
\caption{Voigt Profile Fits for Metal Lines Towards Q1331+17.  Velocities
are relative to $z=0.7450$. }
\vspace*{5mm}
\begin{tabular}{ccccccc}\hline \hline
Cloud &Redshift &  Velocity & $b$ &
\multicolumn{3}{c}{Log$_{10}$ N(X)}\\ 
      &              &  km/s  & km/s & Mg~II & Fe~II   & Mg~I \\ \hline
1     &  0.74385     & $-$197 & 11.0 & 12.11 &  ...    & ...  \\ 
2     &  0.74395     & $-$181 &  3.1 & 13.17 & 12.62  & 10.61 \\
3     &  0.74400     & $-$173 &  5.0 & 12.85 & 12.32  & 10.52 \\
4     &  0.74413     & $-$149 &  4.6 & 13.30 & 12.66  & 10.41 \\
5     &  0.74425     & $-$130 &  4.1 & 12.67 & 12.20  & 10.12 \\
6     &  0.74439     & $-$105 &  7.1 & 13.00 & 12.51  & 10.31 \\
7     &  0.74446     & $-$93  &  3.1 & 12.91 & 12.36  & 11.08 \\
8     &  0.74451     & $-$84  &  4.6 & 12.70 & 12.26  & 10.93 \\
9     &  0.74462     & $-$66  &  3.6 & 14.08 & 13.29  & 11.84 \\
10    &  0.74467     & $-$57  & 10.2 & 13.01 & 12.62  & 10.43 \\
11    &  0.74487     & $-$22  &  4.4 & 12.03 & 11.76  & 10.19 \\
12    &  0.74501     & 2      &  4.0 & 12.21 & 11.88  & 10.50 \\
13    &  0.74522     & 46     & 16.8 & 12.67 & 12.32  & 10.98 \\
14    &  0.74527     & 37     & 10.2 & 12.94 & 12.51  & 10.65 \\
15    &  0.74536     & 61     &  5.2 & 13.08 & 12.45  & 11.04 \\
16    &  0.74550     & 86     & 11.5 & 13.20 & 12.81  & 11.15 \\
17    &  0.74556     & 95    &  2.4 & 14.07 & 13.02  & 11.24 \\ \hline
\multicolumn{3}{c}{Total system N(X)} & & $>$14.56 & 13.84$\pm0.06$ &
12.22$\pm0.05$ \\ \hline
\end{tabular}
\label{vp_fits}
\end{center}
\end{table*}

%
%

\end{document}